\documentclass[10pt,a4paper]{article}
\usepackage[english]{babel}
\usepackage[latin1]{inputenc}
\usepackage{amsfonts,amsbsy,bm,euscript,mathrsfs}
\usepackage{amssymb,stmaryrd,faktor}
\usepackage[tbtags]{amsmath}
\usepackage{bbm}
\usepackage{graphicx}
\usepackage[title,titletoc]{appendix}
\usepackage[bookmarks=true,colorlinks=true,linkcolor=blue,citecolor=blue,urlcolor=blue,bookmarksnumbered]{hyperref}

\usepackage[nosort]{cite}

\newcommand{\AdS}{\textup{AdS}}
\newcommand{\CFT}{\textup{CFT}}
\newcommand{\Sphere}{\textup{S}}
\newcommand{\Torus}{\textup{T}}

\newcommand{\HL}{\mbox{\scriptsize HL}}
\newcommand{\BES}{\mbox{\scriptsize BES}}
\newcommand{\AFS}{\mbox{\scriptsize AFS}}

\numberwithin{equation}{section}

\makeatletter
\renewcommand\section{\@startsection {section}{1}{\z@}
{-3.5ex \@plus -1ex \@minus -.2ex}
{2.3ex \@plus.2ex}
{\normalfont\Large\bfseries}}
\renewcommand\subsection{\@startsection{subsection}{2}{\z@}
{-3.25ex\@plus -1ex \@minus -.2ex}
{1.5ex \@plus.2ex}
{\normalfont\large\bfseries}}
\makeatother

\newcommand{\alg}[1]{\mathfrak{#1}}

\newcommand{\beq}{\begin{equation}}
\newcommand{\eeq}{\end{equation}}

\begin{document}

\thispagestyle{empty}
\begin{flushright}\footnotesize\ttfamily
DMUS-MP-19/09 \\
IFT-UAM/CSIC-19-57 \\
NORDITA 2019-043
\end{flushright}
\vspace{2em}

\begin{center}

{\Large\bf \vspace{0.2cm}
{\color{black} The Effectiveness of Relativistic Invariance in $AdS_3$ }} 
\vspace{1.5cm}

\textrm{\large Andrea Fontanella ${}^{\circ}$\footnote{\texttt{andrea.fontanella@inv.uam.es}}, \ \ \ \ Olof Ohlsson Sax${}^{+}$\footnote{\texttt{olof.ohlsson.sax@nordita.org}}, \\  Bogdan Stefa\'nski, jr.${}^{\bullet,*}$\footnote{\texttt{Bogdan.Stefanski.1@city.ac.uk}}, \ \ \ \ Alessandro Torrielli${}^{\dagger}$\footnote{\texttt{a.torrielli@surrey.ac.uk}}}

\vspace{2em}

\vspace{1em}
\begingroup\itshape
${}^{\circ}$Instituto de F\'isica Te\'orica UAM/CSIC 
\\
C/ Nicol\'as Cabrera, 13-15, C.U. Cantoblanco, 
E-28049 Madrid, Spain
\vspace{0.2cm}
\\
${}^{+}$Nordita, Stockholm University and KTH Royal Institute of Technology
\\
Roslagstullsbacken~23, 106~91 Stockholm, Sweden
\vspace{0.2cm}
\\
${}^{\bullet}$ Perimeter Institute for Theoretical Physics
\\
 31 Caroline Street N, Waterloo, ON N2L 2Y5, Canada
 \vspace{0.2cm}
 \\
${}^{*}$ Centre for Mathematical Science, City, University of London
\\
Northampton Square, EC1V 0HB London, UK
\vspace{0.2cm}
\\
${}^{\dagger}$ Department of Mathematics, University of Surrey
\\
 Guildford, GU2 7XH, UK
\par\endgroup

\end{center}

\vspace{2em}

\begin{abstract}\noindent 
We use relativistic invariance to investigate two aspects of integrable $\AdS_3$ string theory. Firstly, we write down the all-loop TBA equations for the massless sector of the theory with R-R flux, using the recently discovered hidden relativistic symmetry. Secondly, for the low-energy relativistic limit of the theory with NS-NS flux we write down the S matrix, dressing factors and TBA. We find that the integrable system coincides with a restriction to $\AdS_3$ of the relativistic q-deformed $\AdS_5$ theory. We also comment on the relativistic limit of the small-$k$ NS-NS theory.
\end{abstract}

\newpage

\overfullrule=0pt
\parskip=2pt
\parindent=12pt
\headheight=0.0in \headsep=0.0in \topmargin=0.0in \oddsidemargin=0in

\vspace{-3cm}
\thispagestyle{empty}
\vspace{-1cm}

\tableofcontents

\setcounter{footnote}{0}

\section{\label{sec:level1}Introduction}

Relativistic invariance plays an important role in 1+1d integrable field theories. For example, being able to express the S matrix in terms of a single parameter $\theta$ corresponding to the \textit{difference} of rapidities of the scattering particles substantially simplifies the analysis of the system.

By contrast, integrable holographic theories are obtained from a light-cone gauge fixed Green-Schwarz worldsheet action and consequently describe scattering in a non-relativistic 1+1-dimensional theory \cite{Niklas,Gleb}. In $\AdS_3$ integrable models \cite{Bogdan,Cagnazzo:2012se} (see also
\cite{DS1,seealso0,seealso2,seealso1,seealso7,seealso3,seealso6,seealso4,Sundin:2013ypa,s1,
seealso5,s2,seealso8,Babichenko:2014yaa,seealso9,Borsato:2015mma,seealso11,Baggio:2017kza,seealso12}) the exact dispersion relation for massive and massless excitations is~\cite{ArkadyBenStepanchuk,Lloyd:2014bsa}
\begin{equation}
\label{eq:disp-rel}
E = \sqrt{\Big(m + \frac{k}{2 \pi} p\Big)^2 + 4 h^2 \sin^2\frac{p}{2}}\,.
\end{equation}
Here, $k\in\bf{Z}$ is the WZW level, $m$ is the mass of  worldsheet excitations~\footnote{In the case of strings on $\AdS_3\times\Sphere^3\times\Torus^4$ the mass is $m=\pm 1, 0$.}, and $h$ is the strength of the integrable interaction. In the spacetime supported by R-R charge only ($k=0$), $h$ is related to the curvature of the geometry as
\begin{equation}
h= \frac{R^2}{\alpha'}+\dots
\end{equation}
while, for the spacetime supported by NS-NS charge, $h$ corresponds to turning on  R-R moduli~\cite{OhlssonSax:2018hgc}, such as the axion $C_0$
\begin{equation}
h=-\frac{g_sC_0 k}{2\pi}+\dots\,.
\end{equation}
Turning on $C_0$ in the NS-NS theory introduces a non-zero R-R three-form \textit{flux}~\cite{OhlssonSax:2018hgc}, since the gauge-invariant R-R field-strength is
\begin{equation}
F_3=dC_2-C_0\wedge H\,.
\end{equation}
Nonetheless, the \textit{charges} of the background remains the same ($Q_{\mbox{\tiny F1}}$ and
$Q_{\mbox{\tiny NS5}}$), as is expected for a modulus. Because of this we will refer to the theory with non-zero $C_0$ as the NS-NS theory.

 In this paper we exploit the appearance of relativistic invariance in integrable $\AdS_3$ holography in two complementary contexts. Firstly, in the R-R theory ($k=0$), it was recently found in \cite{Fontanella:2019baq} that the massless S matrix can be written in a \textit{difference} form when expressed in terms of unconventional rapidity variables
\begin{equation}
\gamma=\log\tan\frac{p}{4}\,.
\end{equation}
We prove analytically that the dressing factor found in~\cite{Borsato:2016xns} is of $\gamma$-difference form as well. Exploiting this simplification, we write down an exact TBA for the massless degrees of freedom generalising in a straightforward manner the TBA found in~\cite{Bombardelli:2018jkj} for the low-energy excitations. We use the TBA to show that the vacuum remains supersymmetric in finite volume. The S matrix and TBA correspond to a \textit{target-space} supersymmetric massless sine-Gordon theory with a non-conventional kinetic term.~\footnote{The S matrix is almost that of the massless ${\cal N}=2$ sine-Gordon model at $\beta^2_{{\cal N}=2}=16\pi$. The only difference comes from certain constant phases since the fermions considered here are of the GS type, while the usual ${\cal N}=2$ sine-Gordon model has RNS fermions. The former are anti-commuting scalars on the worldsheet, while the latter are spinors.}

Secondly, we investigate the low-energy limit of the NS-NS theory. At zero modulus ($h=0$), the dispersion relation~\eqref{eq:disp-rel} reduces to a relativistic one upon shifting the momentum
\begin{equation}
\label{eq:mom-shift-massive}
p_s= p- \frac{2\pi m}{k}\,.
\end{equation}
The S matrix, however remains non-relativistic. By restricting to small momenta $p_s\sim 0$, we investigate the low-energy limit of the integrable theory, generalising the R-R analysis of~\cite{Bombardelli:2018jkj}. As in that case,  each excitation decomposes into a worldsheet left- and right-mover, the S matrices become relativistic and the only non-trivial scattering is between excitations of the same worldsheet chirality. As was argued on general grounds in~\cite{Zamol2}, the trivialisation of  left/right scattering is expected for S matrices corresponding to (worldsheet) Conformal Field Theories (CFTs). This should be contrasted with~\cite{Baggio,Dei}, where the S matrix is assumed to be trivial apart from a CDD dressing factor for scattering of  excitations of \textit{opposite} worldsheet chirality. It would be interesting to see whether this conjectured S matrix is somehow related to the pure NS-NS limit of the exact all-loop S matrix derived in~\cite{Lloyd:2014bsa}, and whether it can be reconciled with the general expectations for a $\CFT$ S  matrix discussed in~\cite{Zamol2}. 

We find that our low-energy S matrix coincides, up to dressing factors, with a restriction to $\AdS_3$ of the relativistic q-deformed S matrices that have appeared in investigations of $q$-deformations of integrable $\AdS_5$ models~\cite{Relative,Hoare:2011wr} and their Pohlmeyer reductions~\cite{Hoare:2011fj}. The work of \cite{Hoare:2011wr} is concerned with the $\AdS_5\times \Sphere^5$ theory. Restricting to $\AdS_3\times \Sphere^3$ can be done in a straightforward way, and it is such restricted S matrices that we match to our low-energy NS-NS S matrices. Even more closely related are the papers \cite{Hoare:2011fj,Ben} which study the Pohlmeyer reduction directly in $AdS_3 \times S^3$ and $AdS_2 \times S^2$ (see also\cite{Ben,Vidas} for related work in $AdS_3$). The matching with the corresponding S-matrices is reproduced with the only difference being in the scalar factor, accounting for the differences in the physical spectrum of the two cases\footnote{We thank Ben Hoare for communication on this point, and the anonymous referee for a detailed comment about these comparisons.}. This connection gives a new string-theoretic interpretation on the role of $q$-deformations in holographic integrability with $q$ being related to the WZW level of $\AdS_3\times \Sphere^3$ by
\begin{equation}
q = e^{\frac{2 i \pi}{k}}\,.
\end{equation}
We determine the (two) dressing factors of the low-energy S matrices, write down the TBA and show that the vacuum remains supersymmetric in finite volume. Finally, in light of recent interest in the small-$k$ limit of $\AdS_3$ geometries~\cite{Giribet:2018ada,Gaberdiel:2018rqv,Eberhardt:2018ouy}, we consider the relativistic limit of the $k\le 2$ theories. We find that for $k=1,2$ the low-momentum S matrices reduce to the well-known minimal ${\cal{N}}=2$ S-matrix of Fendley-Intriligator \cite{FI}, equivalently the Sine-Gordon S-matrix at the special value of the coupling  $\beta^2 = \frac{16 \pi}{3}$. This suggests the possibility of novel behaviour emerging from the low-$k$ integrable system.

This paper is organised as follows. In section~\ref{sec2}, we analytically prove that the dressing factor found in \cite{Borsato:2016xns} is of difference-form, and we analyse which implications the hidden relativistic invariance has for a putative dressing factor. We write down the all-loop TBA for the massless modes of the R-R theory and use it to show that the correction to the BMN groundstate energy is zero. In section~\ref{sec:Relativistic} we write down the low energy
S matrices of the NS-NS theory and relate them to relativistic $q$-deformed S matrices~\cite{Hoare:2011wr, Hoare:2011fj}. In section~\ref{sec4} we solve the crossing equations for the two dressing factors  of the low-energy theory, obtain the Algebraic Bethe Ansatz (ABA) in section~\ref{sec5} and write down the groundstate TBA equations  in section~\ref{sec6}. In section~\ref{sec7} we comment on the relativistic limit for the $k=1,2$ integrable theories.

\section{Exact massless sector of R-R theory}
\label{sec2}

In this section we use the hidden relativistic invariance found in~\cite{Fontanella:2019baq} to analyse the exact dressing factor, ABA and TBA for the massless sector of the R-R $\AdS_3$ theory.~\footnote{We focus on the $\AdS_3\times\Sphere^3\times\Torus^4$ background, but generalisation to the massless sector of $\AdS_3\times\Sphere^3\times\Sphere^3\times\Sphere^1$ is straightforward.}

\subsection{Proof of difference-form of massless dressing factor}

In this section we give a full proof of the fact discovered in \cite{Fontanella:2019baq} that the non-relativistic massless sector of string theory on $AdS_3 \times S^3 \times T^4$ can be entirely recast into a difference-form in terms of the new variable
\begin{equation}
\gamma = \log \tan \frac{p}{4}.
\end{equation}
Equation (2.33) of~\cite{Borsato:2016xns} gives the massless dressing phase $\theta^{\circ \circ}(x,y)$
\begin{equation}\label{eq:hl-massless}
\begin{aligned}
\theta^{\circ\circ}_{\mbox{\scriptsize min}}(x,y)&=\frac{1}{2}
\theta^{\mbox{\scriptsize HL}} (x^\pm,y^\pm)\Big|_{m_x=m_y=0}
\\
&= \int\limits_{-1+i0}^{1+i0} \frac{dz}{4\pi}G_-(z,y)
\left( \frac{1}{z-x}-\frac{1}{z-\tfrac{1}{x}} \right)
- \int\limits_{-1-i0}^{1-i0} \frac{dz}{4\pi}G_+(z,\tfrac{1}{y})
\left( \frac{1}{z-x}-\frac{1}{z-\tfrac{1}{x}} \right)\\
& \,\,\,\,\,\,-\frac{i}{4} \left( G_-(x,y)-G_+(\tfrac{1}{x},\tfrac{1}{y}) \right)\,,
\end{aligned}
\end{equation}
where
\begin{equation}
G_\pm(z,y)=\log\left(\pm i(y-z)\right)-\log\left(\pm i(y-\tfrac{1}{z})\right)\,.
\end{equation}
The above expression is obtained by analytically continuing the Hernandez-Lopez phase~\cite{Hernandez:2006tk} to zero mass and deforming the contour of integration~\cite{Dorey:2007xn} from the unit circle to the above interval, taking care to avoid the log branch-cuts. Even though it is not immediately manifest, the phase is in fact antisymmetric:
\begin{equation}
\theta^{\circ \circ}(x,y) =- \theta^{\circ \circ}(y,x)\,,
\end{equation}
as proved in Appendix B of~\cite{Borsato:2016xns}. Given that 
\begin{eqnarray}
\partial_{\gamma_1} = \frac{1}{2} (x^2-1)\partial_x\,, 
\end{eqnarray}
for massless Zhukovsky variable $x=e^{ip_1/2}$, and similarly for $y= e^{ip_2/2}$, we want to show that
\begin{equation}
\label{equa}
(x^2  -1) \partial_x \theta^{\circ \circ}(x,y) = - (y^2  -1) \partial_y \theta^{\circ \circ}(x,y)\,.
\end{equation}
To do this, we will use the explicit form of the dressing phase when computing the r.h.s. of (\ref{equa}), and for the l.h.s we will use antisymmetry. 

Note first that 
\begin{equation}
\partial_y \Big[G_-(x,y) - G_+\Big(\frac{1}{x}, \frac{1}{y}\Big)\Big] = 0,  \qquad \partial_x \Big[G_-(y,x) - G_+\Big(\frac{1}{y}, \frac{1}{x}\Big)\Big] = 0,
\end{equation}
so that the terms on the last line of~\eqref{eq:hl-massless} do not contribute. Differentiating the remaining terms on the r.h.s. of~\eqref{eq:hl-massless} w.r.t. $y$ gives a simple integral of a \textit{rational} function
\begin{equation}
\partial_y G_-(z,y) = - \partial_y G_+\Big(z,\frac{1}{y}\Big) = \frac{z^2 - 1}{(y-z)(yz - 1)}.
\end{equation}
The two integrands after the action of $\partial_y$ are therefore the same, and we can take the integrals in (\ref{eq:hl-massless}) to the segment $[-1,1]$ without any danger. All in all, we get
\begin{equation}
(y^2 - 1)\partial_y \theta^{\circ\circ}(x,y) = \frac{1}{2\pi} \int_{-1}^1 dz \frac{(z^2-1)(y^2 - 1)(x^2-1)}{(yz-1)(xz-1)(z-y)(x-z)}.
\end{equation}
If we now compute $(x^2 - 1)\partial_x \theta^{\circ\circ}(x,y)$ by switching $x \leftrightarrow y$ on the r.h.s. of equation~\eqref{eq:hl-massless} and changing sign, then an almost identical calculation produces
\begin{equation}
(x^2 - 1)\partial_x \theta^{\circ\circ}(x,y) = - \frac{1}{2\pi} \int_{-1}^1 dz \frac{(z^2-1)(y^2 - 1)(x^2-1)}{(yz-1)(xz-1)(z-y)(x-z)},
\end{equation}
which completes the proof.

\subsection{Constraints on dressing factor from hidden relativistic invariance}

The massless dressing phase $\theta^{\circ\circ}_{\mbox{\scriptsize min}}$ is meant to be exact in $\alpha'$, yet was derived from the Hernandez-Lopez order (\textit{i.e.} one-loop in $\alpha'$) of the massive dressing phase, perhaps giving one some pause.~\footnote{We are grateful to Sergey Frolov and Tristan McLoughlin for asking us important questions about the massless dressing phase which led us to revisit it in light of the hidden relativistic invariance.} Nonetheless, it was shown in~\cite{Borsato:2016xns} that this is because higher-order terms in the strong-coupling expansion of the DHM expression for the BES phase trivialize for massless kinematics. At the AFS order, however, the dressing phase remains non-trivial 
\begin{equation}
\label{eq:afs-summed}
\chi^{\AFS}(x,y)=\tfrac{1}{x} - \tfrac{1}{y} + (y+\tfrac{1}{y}-x-\tfrac{1}{x})
\log\left(1 - \tfrac{1}{x y}\right)\,,
\end{equation}
but solves a homogeneous crossing equation, and so could be thought of as a CDD-type factor. In~\cite{Borsato:2016xns}, the possibility of having such a factor in the exact massless dressing phase was not excluded. 

It is easy to see that $\chi^{\AFS}$ is not of $\gamma$-difference form 
\begin{equation}
\Bigl((x^2-1)\partial_x +(y^2-1)\partial_y\Bigr)\chi^{\AFS}(x,y)\neq 0\,.
\end{equation}
Since the exact massless S matrix has to be of difference form~\cite{Fontanella:2019baq}, we see that  one cannot add $\chi^{\AFS}$ as a CDD factor to $\theta^{\circ\circ}_{\mbox{\scriptsize min}}$. 

More generally, requiring such a difference form on the $c_{r,s}$ expansion of dressing phases
\begin{equation}
\label{dressing_expansion}
\chi(x,y)=\sum_{r,s=1}^\infty \frac{c_{r,s}}{x^r y^s}\,,
\end{equation}
implies that the coefficients $c_{r,s}$
\begin{equation}
\label{rel_inv_cond}
(r+1)c_{r+1,s}-(r-1)c_{r-1,s}+(s+1)c_{r,s+1}-(s-1)c_{r,s-1}=0\,.
\end{equation}
It is simple to check that the HL-order coefficients 
\begin{equation}
c^{\HL}=\frac{(-1)^{r+s}-1}{\pi}\frac{1}{r^2-s^2}\,,
\end{equation}
satisfy this constraint, as expected. On the other hand, the AFS coefficients do not, nor do any of the higher-order strong-coupling expansion coefficients, or the weak-coupling expansion ones. In Appendix~\ref{App:BES} we check numerically that the $c^{\BES}_{r,s}$ coefficients which were found in~\cite{Beisert:2006ez} as solutions of the Janik crossing equation~\cite{Janik:2006dc}, 
do not satisfy equation~\eqref{rel_inv_cond} for intermediate values of the coupling either. We conclude that hidden relativistic invariance places strong constraints on the form of any putative massless dressing factors, and excludes all candidates suggested by higher-dimensional integrable holographic examples continued to massless kinematics, apart from the HL-order term $\theta^{\circ\circ}_{\mbox{\scriptsize min}}$. 

A related argument based on  gamma-relativistic invariance can be seen to equally call for the absence of other CDD factors. If, in fact, we were to allow a CDD factor, this would instantly violate the requirement of absence of poles in the physical strip $\mbox{Im}\gamma \in (0,\pi)$. CDD factors are of the form
\begin{equation}
CDD = \prod_i [\alpha_i], \qquad [\alpha_i] \equiv \frac{\sinh \gamma + i \sin \alpha_i}{\sinh \gamma - i \sin \alpha_i}, \qquad \alpha_i \in (0,\pi) \, \, \, \forall \, \, i,
\end{equation}
and are typically utilised to add and subtract poles wherever the bound-state spectrum should require it, which will fix the range of the index $i$ and the value of the parameters $\alpha_i$. The information required to establish the range of $i$ and the specific values of $\alpha_i$ is normally determined by other physical considerations. Since two massless particles cannot form a bound state, no place is left for bound-state poles in the massless S-matrix, and the only CDD factor which would preserve this requirement would be a trivial one~\footnote{Stricly speaking, the CDD factor could add a zero in the physical strip, but we confine ourselves to consider this superfluous, as in our case we are already starting from an S-matrix which does not have any pole to cancel in the physical strip. Adding zeroes in the physical strip via CDD factors does occur for instance in Toda theories (Tim Hollowood, private communication), and it is only by computing physical quantities (\textit{e.g.} by means of the TBA) that one can absolutely pinpoint their necessity. Besides checking the vanishing of the Witten index, later on in the paper we shall also compute the central charge of our relativistic mixed-flux theory in the $k=2$ case and find that it is consistent with general expectations. Although far from conclusive, we take this as a further encouragement that CDD factors should not play a role in our treatment. We thank Tim Hollowood for discussions about this point.}. Such pole analysis of course relies on the spectral properties of a local field theory, and on gamma-relativistic invariance. The only way a CDD factor could avoid this contradiction is by breaking gamma-relativistic invariance, which is shown above to be the case for AFS. We believe this to be an indication of how the full massless sector of the theory actually appears to behave as a local relativistic QFT in the gamma variable, albeit one with a non-standard dispersion relation, which puts a remarkable constraint on its description. The above argument also shows that since both the expression~\eqref{eq:hl-massless} and the well-known sine-Gordon one of Zamolodchikovs~\eqref{disp} are \textit{minimal} solutions of the crossing equation, they must be equal to each other in the physical strip as was already verified numerically in~\cite{Bombardelli:2018jkj}.

\subsection{Algebraic Bethe Ansatz and Bethe Equations}

Using the relativistic, $\gamma$-difference form of the S matrix and dressing factors, we perform the ABA to obtain the Bethe equations and the Thermodynamic Bethe Ansatz (TBA). To a large extent it is possible to simply import the results obtained for the relativistic case in \cite{Bombardelli:2018jkj}, with minimal changes.~\footnote{Our results are for the massless sector. In the non-relativistic regime the full set of scattering matrices is non-trivial, and this includes the scattering between massive and massless modes, which we do not consider in the present paper. Massive modes and the coupling between them and massless modes will have to be included in the all-loop ABA and TBA.}

Following the procedure of Appendix B of~\cite{Bombardelli:2018jkj}, which applies without modification since the R-matrix is formally identical when written in $\gamma$-variables, we obtain the exact massless Bethe equations
\begin{eqnarray}
1&=&\prod_{j=1}^{K_0}\tanh\left(\frac{\beta_{1,k}-\gamma_j}{2}\right)\,,\label{betherel1}\nonumber\\
e^{-iL p_k}&=&(-1)^{K_0-1}\prod_{\stackrel{j=1}{j\neq k}}^{K_0}S^2(\gamma_k-\gamma_j)\prod_{j=1}^{K_1}\coth\left(\frac{\beta_{1,j}-\gamma_{k}}{2}\right)\prod_{j=1}^{K_3}\coth\left(\frac{\beta_{3,j}-\gamma_{k}}{2}\right)
\,,~~~~~~~~\label{betherel2}\nonumber\\
1&=&\prod_{j=1}^{K_0}\tanh\left(\frac{\beta_{3,k}-\gamma_j}{2}\right)\,,\nonumber
\label{betherel3}
\end{eqnarray}
where $\beta$ are the auxiliary Bethe roots, and $S(\gamma)$ is the sine-Gordon scalar factor
\begin{eqnarray}
\label{zamog}
S (\gamma) = \prod_{\ell=1}^\infty \frac{\Gamma^2(\ell - \tau) \, \Gamma(\frac{1}{2} + \ell + \tau) \,\Gamma(- \frac{1}{2} + \ell + \tau)}{\Gamma^2(\ell + \tau) \, \Gamma(\frac{1}{2} + \ell - \tau) \,\Gamma(- \frac{1}{2} + \ell - \tau)},
\end{eqnarray}
with $\tau \equiv \frac{\gamma}{2 \pi i}$. The dressing factor (\ref{zamog}) satisfies the crossing and unitarity conditions
\begin{equation}
S(\gamma)S(\gamma + i \pi) = i \tanh \frac{\gamma}{2}, \qquad S(\gamma)S(-\gamma) =1,
\end{equation}
and is a pure phase for real $\gamma$. The asymptotic spectrum also follows from the relativistic case, with the appropriate replacement of variables\footnote{It would be extremely interesting to revisit perturbation theory for the massless sector of $AdS_3$ away from the BMN limit in the light of $\gamma$-relativistic invariance. In contrast to the strict BMN limit in fact, one can compare with perturbative computations \cite{seealso1,Sundin:2013ypa,unita1,unita2,unita3} as all of the modes have a group velocity which is strictly less than the speed of light. Since several mismatches were found in comparing with perturbative calculations, precisely as regards to the S-matrix and to the dispersion relation of massless modes, it would be interesting to see whether $\gamma$-relativistic invariance is indeed observed in perturbation theory and can play a role in reconciling these differences (Lorenzo Bianchi, private communication).}.

\subsection{Exact massless TBA for R-R theory}

Given that the structure of the spectrum is the same as the relativistic one, and, in particular, the solutions to the auxiliary Bethe equations localise on the same locus in the complex $\gamma$ plane as  in the relativistic variable $\theta$~\cite{Bombardelli:2018jkj}, the TBA analysis performed there continues to hold as long as we update the particle-energy in terms of $\gamma$:
\begin{equation}
\label{disp}
E = \frac{2 h}{\cosh \gamma}.
\end{equation} 
We obtain for the total energy
\begin{equation}
\widetilde E = 2 h \int d\gamma \, \frac{1}{\cosh \gamma} \, \rho_0^r(\gamma),
\end{equation}
\begin{eqnarray}
&&\varepsilon_0=\nu_0(\gamma)-\sum_{n=1,3}\phi*(L_{+n}+L_{-n}),\qquad \varepsilon_{\pm n}=-\phi*L_0,\ n=1,3\,,\label{tba}
\end{eqnarray}
where
\beq
\nu_0 (\gamma) \equiv \frac{2 h R}{\cosh \gamma}\,, \qquad
\varepsilon_A\equiv\log\frac{\rho_A^h}{\rho_A^r}\,,\qquad
L_A\equiv\log(1+e^{-\varepsilon_A})\,,\qquad
\phi(\theta) \equiv \frac{1}{2 \pi \cosh \theta}\,,\nonumber
\eeq
 with $A=0,\pm\, n$. The density and kernel we have introduced maintain the same meaning as in \cite{Bombardelli:2018jkj}. The exact ground-state energy is then obtained from the solution of the TBA as
\begin{equation}
 E_{0}(R)=-\frac{h}{\pi}\int d\gamma \frac{1}{\cosh \gamma} \log(1+e^{-\varepsilon_0(\gamma)})\,, 
 \label{energy}
\end{equation}
Taking into account that one has two massless momentum-carrying roots in the full theory (due to the presence of an additional symmetry, dubbed $\alg{su}(2)_\circ$ in \cite{seealso8,seealso9}, which effectively works as to add a second equal and independent copy of the system of Bethe ansatz equations), we conclude that the total ground state energy is $2E_0(R)$, with $E_0(R)$ given by (\ref{energy}). 

As one can see, besides the one-particle energy, the other main difference with respect to \cite{Bombardelli:2018jkj} is that we do not have the distinction between worldsheet  left- and right-movers 
. 
In the exact non-relativistic case considered here, both left- and right-movers  are combined into one particle type by taking the physical region of momenta to be 
\begin{equation}
p \in (0,2 \pi),
\end{equation}
and write a single TBA for this individual particle. The would-be left and right- branches of the dispersion relation do interact, but this is clearly taken into account by the TBA, which now has the non-relativistic dispersion relation (\ref{disp}). This is in many respects reminiscent of a \textit{massive} TBA, with the physical region precisely encompassing 
\begin{equation}
\gamma \in (-\infty, \infty)
\end{equation}
monotonically, as a massive relativistic rapidity would do.

This analogy is reinforced by noting that we are not in a scale-invariant setting any longer. A relativistic massless dispersion of the type $E = M e^\theta$ would imply that any rescaling
\begin{equation}
E \to \lambda E \qquad \mbox{simply amounts to} \qquad \theta \to \theta + \log \lambda, \nonumber
\end{equation}
hence a purely difference-form R-matrix is insensitive to rescalings, if one can transfer the rescaling onto a shift of rapidity. This is clearly not possible for (\ref{disp}). We do not have therefore a conformal  field theory any longer, and the TBA is significantly more difficult to solve -- in particular, the ground-state energy is expected to depend non-trivially on $R$, and not simply display the $\frac{1}{R}$ dependence dictated by dimensional analysis as it happens for a CFT. The effective central charge itself will be a function of $R$:
\beq
c_{eff}(R)=-\frac{6R}{\pi}E_0(R).
\label{c}
\eeq 
Nonetheless, one can still evaluate the Witten index since the calculation is independent of the different expression for the energy now featuring in the TBA. Adding a chemical potential which corresponds to a shift in the auxiliary pseudo-energies $\epsilon_n \to \epsilon_n \pm i \pi$, the same constant solutions to the TBA described at the end of section (5.3) of~\cite{Bombardelli:2018jkj} are still valid, and the new energy term  does not affect them. The integral is again zero thanks to the evaluation of the integrand on these solutions.\footnote{We remark that we are working with the \textit{direct} theory, as opposed to the \textit{mirror} one familiar from the $AdS_5 \times S^5$ string theory work of \cite{mirror1,mirror2,mirror3} and \cite{TBA0,TBA1,TBA2}. The integration contour and expression for the single-particle energy will change in mirror kinematics, however the value of the integral evaluated with the chemical potentials turned to $\pm i \pi$ should still be zero. We will return to the interesting question of the interplay between  $\gamma$-relativistic invariance and mirror transformation in the TBA in future work.}
 The Witten index is exactly zero, which means that the massless sector on its own preserves supersymmetry of the ground state. This puts a constraint on the massive sector: if supersymmetry is preserved in the ground state of the complete TBA, the massive modes should also contribute exactly zero to it. 

\section{\label{sec:Relativistic}Relativistic limit of mixed flux case}
In this section we study the relativistic limit of the symmetry algebra and S-matrix, which is more involved in the mixed-flux case with respect to the pure R-R case. In particular, thanks to the peculiar momentum-dependence of the dispersion relation 
\begin{equation}
E = \sqrt{\Big(m + \frac{k}{2 \pi} p\Big)^2 + 4 h^2 \sin^2\frac{p}{2}}
\end{equation}
it is possible to obtain a much wider set of massless particles participating in the scattering, even starting from those which are initially massive. The prescription to obtain a \textit{massless} relativistic scattering theory goes as follows:

\begin{itemize}

\item Take the limit $h \to 0$ - this eliminates the $\sin\frac{p}{2}$ dependence of the dispersion relation, and makes it natural to shift momenta to reabsorb a constant;

\item Shift momenta as $p \to p - \frac{2 \pi |m|}{k}$ for  representation $\rho_L$, and $p \to p + \frac{2 \pi |m|}{k}$ for  representation $\rho_R$;

\item Rescale all  momenta as $p_i = \epsilon q_i$ and take the limit $\epsilon \to 0$;

\item Set $q_i = e^{\theta_i}$, with $i=1,2$.
 
\end{itemize}

The small-$h$ expansion of the variables $x^\pm$ depends on the sign of $2\pi\pm k p$, which physically corresponds to the worldsheet chirality of the relativistic particles. For $|m|=1$, one finds
\begin{equation}
  \begin{aligned}
    x_L^+ &\sim +\frac{1+\frac{k}{2\pi}p}{h \sin\frac{p}{2}} e^{+\frac{ip}{2}},  \qquad & x_L^- &\sim +\frac{1+\frac{k}{2\pi}p}{h \sin\frac{p}{2}} e^{-\frac{ip}{2}}, \qquad & 2 \pi +k p & >0,\\
    x_L^+ &\sim -\frac{h \sin\frac{p}{2}}{1+\frac{k}{2\pi}p} e^{+\frac{ip}{2}},  \qquad & x_L^- &\sim -\frac{h \sin\frac{p}{2}}{1+\frac{k}{2\pi}p} e^{-\frac{ip}{2}}, \qquad & 2 \pi +k p & <0,\\
    x_R^+ &\sim +\frac{1-\frac{k}{2\pi}p}{h \sin\frac{p}{2}} e^{+\frac{ip}{2}},  \qquad & x_R^- &\sim +\frac{1-\frac{k}{2\pi}p}{h \sin\frac{p}{2}} e^{-\frac{ip}{2}}, \qquad & 2 \pi -k p & >0,\\
    x_R^+ &\sim -\frac{h \sin\frac{p}{2}}{1-\frac{k}{2\pi}p} e^{+\frac{ip}{2}},  \qquad & x_R^- &\sim -\frac{h \sin\frac{p}{2}}{1-\frac{k}{2\pi}p} e^{-\frac{ip}{2}}, \qquad & 2 \pi -k p & <0,
  \end{aligned}
\end{equation}
 One has to then calculate the limit of the algebra and S-matrix for all possible combinations of representations $\rho_L$ and $\rho_R$ and choices of sign of shifted momenta. In particular this means that each exact non-relativistic representation degenerates into two \textit{distinct} relativistic representations, much as we saw in the R-R theory. As a consistency check, we have also rederived all the relevant S-matrices directly starting from the intertwining equation with the limiting algebra generators, suitably rescaled to maintain their finiteness. 

The limit behaves differently for $m=0$ and for $m\neq 0$. Let us first discuss the case of $|m|=1$.
The limiting S-matrix becomes a relativistic one, depending only on the difference of the massless rapidities $\theta_i$, or, equivalently, only on the ratio $\frac{q_1}{q_2}$. In the case of L-L scattering representations, and for the choice of all positive shifted momenta, the resulting S-matrix is~\footnote{These expressions give a good small-momentum approximation to the exact S-matrix for $k>2$, and in most of the rest of the paper we will only consider $k>2$.}~\footnote{For brevity, in most of the explicit formulas in this paper we only write down one of the two identical $\alg{su}(2)_\circ$ terms that the $\AdS_3\times\Sphere^3\times\Torus^4$ theory has.}
\begin{equation}\label{eq:RLLlim1}
  \begin{aligned}
    R_{LL} |\phi\rangle \otimes |\phi\rangle &= |\phi\rangle \otimes |\phi\rangle, \\
    R_{LL} |\phi\rangle \otimes |\psi\rangle &= \frac{e^{\frac{i \pi}{k}}(q_1 - q_2)}{e^{\frac{2 i \pi}{k}} q_1 - q_2} |\phi\rangle \otimes |\psi\rangle +  \frac{(e^{\frac{2 i \pi}{k}}-1) \sqrt{q_1 q_2}}{e^{\frac{2 i \pi}{k}} q_1 - q_2} |\psi\rangle \otimes |\phi\rangle, \\
    R_{LL}|\psi\rangle \otimes |\phi\rangle &= \frac{(e^{\frac{2 i \pi}{k}}-1) \sqrt{q_1 q_2}}{e^{\frac{2 i \pi}{k}} q_1 - q_2} |\phi\rangle \otimes |\psi\rangle +  \frac{e^{\frac{i \pi}{k}}(q_1 - q_2)}{e^{\frac{2 i \pi}{k}} q_1 - q_2}  |\psi\rangle \otimes |\phi\rangle, \\
    R_{LL} |\psi\rangle \otimes |\psi\rangle &= -\frac{e^{\frac{2 i \pi}{k}} q_2- q_1}{e^{\frac{2 i \pi}{k}} q_1 - q_2}\, |\psi\rangle \otimes |\psi\rangle.
  \end{aligned}
\end{equation}
On the other hand for $m=0$ we find 
\begin{equation}\label{eq:RLLlim0}
  \begin{aligned}
    R_{LL} |\phi\rangle \otimes |\phi\rangle\ &= |\phi\rangle \otimes |\phi\rangle, \\
    R_{LL} |\phi\rangle \otimes |\psi\rangle\ &= - \frac{q_1 - q_2}{q_1 + q_2} |\phi\rangle \otimes |\psi\rangle +  \, \frac{2 \sqrt{q_1 q_2}}{q_1 + q_2} |\psi\rangle \otimes |\phi\rangle, \\
    R_{LL}|\psi\rangle \otimes |\phi\rangle\ &= \frac{2 \sqrt{q_1 q_2}}{q_1 + q_2} |\phi\rangle \otimes |\psi\rangle + \frac{q_1 - q_2}{q_1 + q_2}  |\psi\rangle \otimes |\phi\rangle, \\
    R_{LL} |\psi\rangle \otimes |\psi\rangle\ &= - |\psi\rangle \otimes |\psi\rangle,
  \end{aligned}
\end{equation}
which coincides with the one studied in \cite{Bombardelli:2018jkj} for the relativistic massless sector of the pure R-R-fux background. From the point of view of worldsheet scattering theory, the above S matrices should be interpreted as \textit{non-perturbative, algebraic} objects which provide an integrable description of the low-momentum effective CFT in the sense of~\cite{Zamol2}. 

The algebra of symmetries undergoes an interesting reduction in the relativistic limit. First, we notice how, in the $m=0$ case, the algebra generators become proportional to the one studied in \cite{Bombardelli:2018jkj} by constant factors, and the coproduct trivialises. On the other hand for $|m|=1$, because of the shift in momenta~\eqref{eq:mom-shift-massive} \textit{the braiding of the coproduct survives in the relativistic limit}. The braiding factors reduce to constant matrices proportional to identity. Explicitly, one gets for example supercharges in the $\rho_L$ representation with positive shifted momentum
\begin{eqnarray}
\label{leftreprel}
&&\mathfrak{Q}_{L} \sim e^{- \frac{i \pi}{2 k}+ i \xi} \sqrt{\frac{k \epsilon q}{2 \pi}}\begin{pmatrix}0&0\\1&0\end{pmatrix},\qquad
\qquad \, \, \, \, \mathfrak{S}_{L} \sim e^{\frac{i \pi}{2 k}- i \xi} \sqrt{\frac{k \epsilon q}{2 \pi}}\begin{pmatrix}0&1\\0&0\end{pmatrix},\nonumber\\
&& \mathfrak{Q}_{R} \sim e^{- \frac{i \pi}{2 k}+ i \xi} h \sin \frac{\pi}{k} \, \sqrt{\frac{2 \pi}{k \epsilon q}}\begin{pmatrix}0&1\\0&0\end{pmatrix}\qquad \mathfrak{S}_{R} \sim  h e^{\frac{i \pi}{2 k}- i \xi} \sin \frac{\pi}{k} \sqrt{\frac{2 \pi}{k \epsilon q}}\begin{pmatrix}0&0\\1&0\end{pmatrix},
\end{eqnarray}
and for  supercharges in the $\rho_R$ representation with negative shifted momentum
\begin{eqnarray}
\label{rightreprel}
&&\mathfrak{Q}_{L} \sim i e^{\frac{i \pi}{2 k}+ i \xi} \sqrt{\frac{- k \epsilon q}{2 \pi}}\begin{pmatrix}0&1\\0&0\end{pmatrix},\qquad
\, \, \, \, \qquad \, \, \, \mathfrak{S}_{L} \sim i e^{-\frac{i \pi}{2 k}- i \xi} \sqrt{\frac{-k \epsilon q}{2 \pi}}\begin{pmatrix}0&0\\1&0\end{pmatrix},\nonumber\\
&&\mathfrak{Q}_{R} \sim i e^{\frac{i \pi}{2 k}+ i \xi} h \sin \frac{\pi}{k} \, \sqrt{\frac{2 \pi}{-k \epsilon q}}\begin{pmatrix}0&0\\1&0\end{pmatrix},\qquad\mathfrak{S}_{R} \sim i h e^{-\frac{i \pi}{2 k}- i \xi} \sin \frac{\pi}{k} \, \sqrt{\frac{2 \pi}{- k \epsilon q}}\begin{pmatrix}0&1\\0&0\end{pmatrix},
\end{eqnarray}
with a coproduct which has retained a non-trivial braiding
\begin{eqnarray}
\label{Deltac}
&&\Delta(\mathfrak{Q}_{L}) = (\mathfrak{Q}_{L} \otimes \mathfrak{1} +U \otimes \mathfrak{Q}_{L}), \qquad \Delta(\mathfrak{S}_{L}) = (\mathfrak{S}_{L} \otimes \mathfrak{1} + U^{-1} \otimes \mathfrak{S}_{L}), \nonumber\\
&&\Delta(\mathfrak{Q}_{R}) = (\mathfrak{Q}_{R} \otimes \mathfrak{1} +U \otimes \mathfrak{Q}_{R}), \qquad \Delta(\mathfrak{S}_{R}) = (\mathfrak{S}_{R} \otimes \mathfrak{1} + U^{-1} \otimes \mathfrak{S}_{R}),
\end{eqnarray}
with $U$ an invertible central element of the algebra, equal to $e^{-\frac{i \pi}{k}}$ (respectively $e^{\frac{i \pi}{k}}$) for $\rho_L$ with positive momentum (respectively for $\rho_R$ with negative momentum). This defines a consistent relativistic Hopf algebra with $\Delta^{op} \neq \Delta$.~\footnote{As we had anticipated, the  generators have to be suitably rescaled to remain finite in the relativistic limit. This rescaling does not introduce any ambiguity in the calculation of the R-matrices, since the representations involved have a corresponding dependence on the small parameters, and so the rescaling does not affect the intertwining relation. }

\subsection{Comparison with the $q$-deformed S-matrix}

It is interesting to notice that the $|m|=1$ relativistic S-matrix coincides with the relativistic $q$-deformed S-matrix of \cite{Hoare:2011wr} -- and the closely related S-matrix computed in \cite{Hoare:2011fj} in the context of the Pohlmeyer reduction -- upon the following identification of the parameters:
\begin{equation}
q = e^{\frac{2 i \pi}{k}},
\end{equation}
where $q$ is the parameter used in \cite{Hoare:2011wr}, while $k$ is the WZW level used in this paper. This gives a simple string-theory embedding for these models\footnote{We remark that a series of work, primarily \cite{Timminus1,Tim0} and culminated in \cite{Tim1}, pointed out the appearance of a lack of unitarity in the S-matrix studied in the context of $q$-deformations of $AdS_5 \times S^5$. In \cite{Tim1} it was shown how this could be rectified by a specific \textit{vertex to IRF} transformation. We will construct a particular scattering matrix directly in the relativistic limit and massless kinematics for the particular sub-algebra we are concerned with, which we will be using in the TBA, and which does not seem to suffer from those issues. We shall, in the particular case of $k=2$, show that the matrix we will use in the TBA reduces to the ${\cal{N}}=2$ scattering matrix of Fendley and Intriligator, and we will derive the central charge from the TBA in that case. We also remark that, while the transformation mentioned in \cite{Tim1} does generically affect the TBA, in our case we do not have strings as we do not have bound states, neither in the direct nor in the mirror theory. It would be very interesting to investigate whether our findings indicate a possible lift to a unitary theory parallel to the one traditionally studied in $q$ deformations, at least for the specific subsector we focus on and perhaps restricting to the massless sector. We thank Tim Hollowood for discussion about these issues.}. In this respect, even more closely related is a comparison with the Pohlmeyer Reduction of the $AdS_3 \times S^3$ S-matrix~studied in \cite{Ben,Hoare:2011fj}. The matching there is reproduced with the only difference residing in the scalar factor, which takes into account the differences in the physical spectrum of the two situations\footnote{This particular comparison also highlights that no problems with unitarity are encountered when the parameter $q$ is taken to be a phase}. This underlines how a web of connections is reachable starting from the class of relativistic S-matrices we study in this paper~\footnote{We thank the anonymous referee for a detailed and very helpful explanation of these facts.}. 

\section{Crossing and Braiding Unitarity}
\label{sec4}

In this section we show that the relativistic S matrices found above are unitary and satisfy Hopf-algebra crossing relations. We solve the corresponding crossing equations for the dressing factors of the theory.

\subsection{Crossing}

One starts by imposing the fundamental crossing equation for the algebra generators:
\begin{eqnarray}
\label{4}
&&\rho_{L,pos} \Sigma (a) = C^{-1} [\rho_{R,neg} a]^{st}(\theta + i \pi) C, \qquad a = \mathfrak{Q}_L, \mathfrak{S}_L,\mathfrak{Q}_R,\mathfrak{S}_R,
\end{eqnarray}
where \textit{pos} and \textit{neg} indicate positive and negative shifted momentum, respectively, the apex \textit{st} denotes supertransposition, and $C$ is the charge conjugation matrix. $\Sigma$ is the Hopf algebra antipode, which is uniquely fixed from the knowledge of the coproduct (\ref{Deltac}), and reads 
\begin{eqnarray}
&&\Sigma({\mathfrak{Q}_L}) = - U^{-1}\mathfrak{Q}_L, \qquad \Sigma({\mathfrak{S}_L}) = - U \mathfrak{S}_L, \qquad \Sigma({\mathfrak{Q}_R}) = - U^{-1}\mathfrak{Q}_R, \qquad \Sigma({\mathfrak{S}_R}) = - U\mathfrak{S}_R.
\end{eqnarray}
One finds that the four equations (\ref{4}) are solved for a common
\begin{equation}
C = \mbox{diag}(1,-i).
\end{equation}

This then allows us to derive all the required crossing equations. We remark that \textit{the $\rho_L$ representation correlates with positive shifted momentum, and the $\rho_R$ representation with negative shifted momentum. Both combinations are considered worldsheet right movers.} This stems from considering that the small-$h$ small-momentum behaviour of the supercharges naturally pairs up if we inspect (\ref{leftreprel}) and (\ref{rightreprel}) and allows to build a consistent set of intertwining and crossing equations in the strict relativistic limit, without being upset by different scaling behaviours of the algebra generators.
  
It is then just a matter of evaluating the abstract Hopf-algebra equation
\begin{equation}
(\Sigma \otimes \mathfrak{1}) R = R^{-1}
\end{equation}
in all possible representations. We have first derived all the R-matrices in the appropriate representations by solving the intertwining equation 
\begin{equation}
\Delta^{op} (a) R = R \Delta(a),\qquad \qquad a = \mathfrak{Q}_L, \mathfrak{S}_L,\mathfrak{Q}_R,\mathfrak{S}_R,
\end{equation}
for the coproduct (\ref{Deltac}) in all possibile combinations of representations. Noticed that we have conventionally normalised the R-matrices $R_{LL}$ and $R_{RR}$ such that the entry $|\phi\rangle \otimes |\phi\rangle \to |\phi\rangle \otimes |\phi\rangle$ always equals $\Phi(\theta)$, this being the corresponding dressing factor. Likewise, we have conventionally normalised the R-matrices $R_{RL}$ and $R_{LR}$ such that the entry $|\phi\rangle \otimes |\psi\rangle \to |\phi\rangle \otimes |\psi\rangle$ always equals $\Phi(\theta)$, this being the corresponding dressing factor.

Focusing on the worldsheet right movers, we therefore get the crossing equations:  
\begin{eqnarray}
\label{systema}
&&R_{LL,pos - pos}(\theta) \, C_1^{-1} R_{RL,neg - pos}^{st_1} (\theta + i \pi) C_1 = \frac{1}{f(\theta)}\mathfrak{1} \otimes \mathfrak{1},  \qquad  \mbox{hence} \qquad \Phi_{LL}(\theta) \Phi_{RL}(\theta + i \pi) = f(\theta), \nonumber \\
&&R_{RL,neg - pos}(\theta) \, C_1 R_{LL,pos - pos}^{st_1} (\theta + i \pi) C_1^{-1} = \frac{1}{g(\theta)}\mathfrak{1} \otimes \mathfrak{1}, \qquad  \mbox{hence} \quad \Phi_{RL}(\theta) \Phi_{LL}(\theta + i \pi) = g(\theta), \nonumber \\
\end{eqnarray}
where 
\begin{equation}
\label{fg}
f(\theta) = \frac{\sinh \frac{\theta}{2}}{\sinh (\frac{\theta}{2}- \frac{i \pi}{k})}, \qquad g(\theta) = \frac{\cosh (\frac{\theta}{2}+ \frac{i \pi}{k})}{\cosh \frac{\theta}{2}},
\end{equation}
$st_1$ means supertransposition in the first  factor and $C_1^{\pm 1} \equiv C^{\pm 1} \otimes \mathfrak{1}$.

\subsection{Braiding and Physical Unitarity}

Braiding unitarity is obtained by evaluating the abstract Hopf-algebra equation
\begin{equation}
R_{21} \, R_{12} = \mathfrak{1} \otimes \mathfrak{1}
\end{equation}
in all possible representations. We simply state the result we obtain:
\begin{eqnarray}
\label{systema1}
&&R_{LL,pos - pos}(\theta) \, R^{op}_{LL,pos - pos}(-\theta) = \mathfrak{1} \otimes \mathfrak{1},  \qquad  \mbox{hence} \qquad \Phi_{LL}(\theta) \Phi_{LL}(-\theta) = 1, \nonumber \\
&&R_{RR,neg - neg}(\theta) \, R^{op}_{RR,neg - neg}(-\theta) = \mathfrak{1} \otimes \mathfrak{1},\qquad  \mbox{hence} \qquad \Phi_{RR}(\theta) \Phi_{RR}(-\theta) = 1, \nonumber \\
&&R_{LR,pos - neg}(\theta) \, R^{op}_{RL,neg - pos}(-\theta) =  \mathfrak{1} \otimes \mathfrak{1}, \qquad  \mbox{hence} \qquad \Phi_{LR}(\theta) \Phi_{RL}(-\theta) = 1, \nonumber \\
&&R_{RL,neg - pos}(\theta) \, R^{op}_{LR,pos - neg}(-\theta) =  \mathfrak{1} \otimes \mathfrak{1},  \qquad  \mbox{hence} \qquad \Phi_{RL}(\theta) \Phi_{LR}(-\theta) = 1.
\end{eqnarray}

All the equations (\ref{systema}) and (\ref{systema1}) will need to be solved simultaneously to determine the dressing factors to insert in the TBA.

Finally, we mention physical unitarity - starting with
\begin{eqnarray}
&&R_{LL,pos-pos}^\dagger (\theta) \, R_{LL,pos-pos} (\theta) = \mathfrak{1} \otimes \mathfrak{1},\quad R_{RR,neg-neg}^\dagger (\theta) \, R_{RR,neg-neg} (\theta) = \mathfrak{1} \otimes \mathfrak{1},
\end{eqnarray}
for $\theta$ real. Given that the matrix parts, (\ref{eq:RLLlim1}) and $RR$-analog, turn out to be unitary by itself for real $\theta$, we shall need to impose for instance
\begin{equation}
|\Phi_{LL}(\theta)| = 1, \qquad |\Phi_{RR}(\theta)| = 1,
\end{equation}
for real $\theta$. By carefully resolving some ambiguities in the conjugation of square-root factors, one can also show
\begin{eqnarray}
&&R_{RL,pos-neg} (-\theta) \, \Big[R_{LR,neg-pos}^\dagger (\theta)\Big]^{-1} = \mathfrak{1} \otimes \mathfrak{1},\quad R_{LR,neg-pos} (-\theta) \, \Big[R_{RL,pos-neg}^\dagger (\theta)\Big]^{-1} = \mathfrak{1} \otimes \mathfrak{1},\nonumber
\end{eqnarray}  
for the matrix parts themselves for real $\theta$, which in turn implies that the dressing factors have to satisfy 
\begin{equation}
\label{mixu}
\frac{\Phi_{RL}(-\theta)}{\Phi_{LR}^*(\theta)}=1, \qquad \frac{\Phi_{LR}(-\theta)}{\Phi_{RL}^*(\theta)}=1, 
\end{equation}
for real $\theta$. Using the conditions on the dressing factors descending from braiding unitarity (\ref{systema1}), one can see that the mixed $L-R$ physical unitarity conditions (\ref{mixu})  reduce to the corresponding dressing factors being pure phases for real momenta.

In later chapters we shall focus on $R_{LL}$ (and omit the $pos-pos$ index) for the Bethe ansatz and the TBA. The R-matrix $R_{LL}$ satisfies in addition the generalised physical unitarity condition for any $\theta$
\begin{equation}
R_{LL}^\dagger (\theta^*) \, R_{LL}(\theta)= \mathfrak{1} \otimes \mathfrak{1}.
\end{equation}

\subsection{Dressing factors}



By adopting the same procedure which was employed in \cite{Borsato:2013hoa}, we single out two dressing factors. Eventually, $L\leftrightarrow R$ symmetry might be invoked to obtain the other two. If we focus on $\Phi_{LL}$ and $\Phi_{LR}$, we see that we can define two combinations
\begin{equation}
\sigma_+ (\theta) \equiv \Phi_{LL}(\theta) \Phi_{RL}(\theta), \qquad \sigma_- (\theta) \equiv \frac{\Phi_{LL}(\theta)}{\Phi_{RL}(\theta)}, 
\end{equation}
which satisfy two decoupled crossing equations:
\begin{eqnarray}
&&\label{1}\sigma_+(\theta) \, \sigma_+(\theta + i \pi) = f(\theta)g(\theta), \\
&& \label{2}\frac{\sigma_-(\theta)}{\sigma_-(\theta + i \pi)} = \frac{f(\theta)}{g(\theta)} = \frac{\sinh \theta}{\sinh \theta - i \sin \frac{2 \pi}{k}}.
\end{eqnarray} 
We can notice that the function appearing on the r.h.s. of the crossing equation for $\sigma_+$ is given by 
\begin{equation}
f(\theta) g(\theta) = \frac{\sinh \frac{\theta}{2} \, \, \cosh(\frac{\theta}{2} + \frac{i \pi}{k})}{\cosh \frac{\theta}{2} \, \, \sinh(\frac{\theta}{2} - \frac{i \pi}{k})}. 
\end{equation}
We can compare this with the dressing phase (5.41) and (5.42) in \cite{Hoare:2011wr}, which satisfies
\begin{equation}
\sigma_{hhm}(\theta) \, \sigma_{hhm}(\theta + i \pi) = \frac{\sinh \frac{\theta}{2} \, \, \cosh(\frac{\theta}{2} + \frac{i \pi}{2 k_{hhm}})}{\cosh \frac{\theta}{2} \, \, \sinh(\frac{\theta}{2} - \frac{i \pi}{2 k_{hhm}})},
\end{equation}
$\sigma_{hhm}(\theta)$ and $k_{hhm}$ respectively being what are called $\sigma(\theta)|_{\mbox{magnon}}$ and $k$ in (5.41) and (5.42) of \cite{Hoare:2011wr}. Comparing with (\ref{1}), we see that we can borrow the solution provided in \cite{Hoare:2011wr}, which is meromorphic in the entire complex $\theta$-plane. We report the explicit expression here below:
\begin{eqnarray}
\label{sigp}
&&\sigma_+(\theta) = \frac{\cosh (\frac{\theta}{2} + \frac{i \pi}{k})}{\cosh (\frac{\theta}{2} - \frac{i \pi}{k})} 
\prod_{\ell=0}^\infty \frac{\Gamma(\tau + \frac{1}{2} +\ell)}{\Gamma(-\tau + \frac{1}{2} +\ell)}
\frac{\Gamma(\tau + \frac{1}{k}+1 +\ell)}{\Gamma(-\tau + \frac{1}{k}+1 +\ell)}  
\frac{\Gamma(-\tau +1 +\ell)}{\Gamma(\tau +1 +\ell)} 
\frac{\Gamma(-\tau - \frac{1}{k} + \frac{1}{2} + \ell)}{\Gamma(\tau - \frac{1}{k} + \frac{1}{2} + \ell)}\nonumber\\
&& \qquad \qquad \frac{\Gamma(-\tau + 1 +\ell)}{\Gamma(\tau + 1 +\ell)}
\frac{\Gamma(-\tau + \frac{1}{k} + \frac{3}{2} + \ell)}{\Gamma(\tau + \frac{1}{k} + \frac{3}{2} + \ell)} 
\frac{\Gamma(\tau +\frac{3}{2}+\ell)}{\Gamma(-\tau +\frac{3}{2}+\ell)}
\frac{\Gamma(\tau -\frac{1}{k}+1+\ell)}{\Gamma(-\tau -\frac{1}{k}+1+\ell)},
\end{eqnarray}
where we have defined
\begin{equation}
\tau \equiv \frac{\theta}{2 \pi i}.
\end{equation}
We have verified by hand, using the properties of Gamma functions under integer shifts, and the product representation of $\sin$ and $\cos$, that  (\ref{sigp}) satisfies (\ref{1}). We can also notice that, for real $\theta$, $\sigma_+$ given by (\ref{sigp}) is a pure phase.

We can provide a solution to the crossing equation for $\sigma_-$ by making use of Fourier transforms. If we define
\begin{equation}
\tilde{F}(\omega) \equiv \frac{1}{\sqrt{2\pi}} \int_{-\infty}^\infty d\theta e^{i \omega \theta} \, \log \sigma_-(\theta),
\end{equation}
we can use the properties of the Fourier transform under shift of argument to re-write the (logarithmic version of) crossing equation (\ref{2}) as
\begin{equation}
\tilde{F}(\omega) - e^{\omega \pi}\tilde{F}(\omega) = \tilde{G}(\omega),
\end{equation}
where $\tilde{G}(\omega)$ is the Fourier transform of the logarithm of the r.h.s. of (\ref{2}):
\begin{equation}
\tilde{G}(\omega) \equiv \frac{1}{\sqrt{2\pi}} \int_{-\infty}^\infty d\theta e^{i \omega \theta} \, \log \frac{\sinh \theta}{\sinh \theta - i \sin \frac{2 \pi}{k}}.
\end{equation}
In the region $\mbox{Im}[\omega] \in (-1,0)$ one has for example for $k=4$
\begin{eqnarray}
&&\tilde{G}(\omega)=\frac{1}{\sqrt{2 \pi} \, \omega^2}\Big[2 - \pi \omega - \pi \omega \, \mbox{coth} \frac{\pi \omega}{2}+ 2 i \omega e^{-\frac{\pi \omega}{2}} \Big(B(-i,1+i\omega,0)-B(i,1-i\omega,0)\Big)\Big],\nonumber
\end{eqnarray}
where $B(a,b,c)$ is the \textit{incomplete Beta function}.

We can therefore immediately solve the crossing equation as \begin{equation}
\tilde{F}(\omega) = \frac{\tilde{G}(\omega)}{1-e^{\omega \pi}},
\end{equation}
hence
\begin{equation}
\sigma_-(\theta) = \exp \frac{1}{\sqrt{2\pi}} \int_{-\infty}^\infty d\omega e^{-i \omega \theta} \, \frac{\tilde{G}(\omega)}{1-e^{\omega \pi}}.
\end{equation}

To improve convergency at $\omega=0$, one might want to evaluate
\begin{equation}
\label{32}
\frac{d^m \log \sigma_-(\theta)}{d\theta^m} = \frac{1}{\sqrt{2\pi}} \int_{-\infty}^\infty d\omega (-i \omega)^m e^{-i \omega \theta} \, \frac{\tilde{G}(\omega)}{1-e^{\omega \pi}}
\end{equation}
for some natural number $m$. If we consider our example of $k=4$, since the incomplete Beta function is analytic in all its three arguments, we conclude that, for an appropriate value of $m$, the integrand of (\ref{32}) has no poles on the real axis, and is meromorphic in the $\omega$ complex plane with all the potential poles located along the imaginary axis. 


It is interesting to note that upon setting $k=2$ we recover the Fendley-Intriligator (\textit{minimal ${\cal{N}}=2$}) supersymmetric scattering theory~\cite{FI}. One can see that $\sigma_- = 1$ in this case, and one can check directly that $\sigma_+$ reduces for $k=2$ to the square of $S(\theta)$, the famous Zamolodchikov phase factor for Sine-Gordon at the same special value of the coupling $\beta^2 = \frac{16 \pi}{3}$, namely
\begin{equation}
\label{zamoga}
S (\theta) = \prod_{\ell=1}^\infty \frac{\Gamma^2(\ell - \tau) \, \Gamma(\frac{1}{2} + \ell + \tau) \,\Gamma(- \frac{1}{2} + \ell + \tau)}{\Gamma^2(\ell + \tau) \, \Gamma(\frac{1}{2} + \ell - \tau) \,\Gamma(- \frac{1}{2} + \ell - \tau)},
\end{equation}
as for the massless sector. We will analyse the consequences of this observation in the future.

\section{Algebraic Bethe Ansatz}
\label{sec5}

In this section we construct the ABA corresponding to the R-matrix (\ref{eq:RLLlim1}). We will follow the treatment in Appendix B of \cite{Bombardelli:2018jkj} and use the same conventions adopted there. The Bethe ansatz can be performed explicitly along similar lines as in the pure R-R case. 
We begin by writing the R-matrix (\ref{eq:RLLlim1}) as 
\begin{eqnarray}
&&R_{LL}(\theta) = E_{11} \otimes E_{11} + c(\theta) E_{22} \otimes E_{22} + b(\theta) (E_{11} \otimes E_{22}+ E_{22} \otimes E_{11}) +a(\theta) (E_{21} \otimes E_{12}- E_{12} \otimes E_{21}), \nonumber
\end{eqnarray}
with the assignement
\begin{eqnarray}
\label{abc}
&&a(\theta) = \frac{e^{\frac{\theta}{2}} (e^{\frac{2 \pi i }{k}} - 1)}{e^{\frac{2 \pi i }{k}+\theta} - 1}, \qquad b(\theta) = \frac{e^{\frac{\pi i }{k}} (e^\theta - 1)}{e^{\frac{2 \pi i }{k}+\theta} - 1}, \qquad
c(\theta) = \frac{e^\theta - e^{\frac{2 \pi i }{k}}}{e^{\frac{2 \pi i }{k}+\theta} - 1},
\end{eqnarray}
where we have parameterised as usual $q_i = e^{\theta_i}$, $i=1,2$, and set $\theta = \theta_1 - \theta_2$.

As in \cite{Bombardelli:2018jkj}, we can use the pseudo-vacuum 
\begin{eqnarray}
|0\rangle = |\phi \rangle \otimes ... \otimes |\phi \rangle
\end{eqnarray}
to implement the algebraic Bethe ansatz method. The central object is the transfer matrix 
\begin{eqnarray}
{\cal{T}}(\theta_0|\vec{\theta}\, ) = str_0 {\cal{M}}(\theta_0|\vec{\theta}\, ), \, {\cal{M}}(\theta_0|\vec{\theta}\, ) = \prod_{i=1}^N R_{0i}(\theta_0 - \theta_i),\nonumber
\end{eqnarray}
${\cal{M}}(\theta_0|\vec{\theta}\, )$ being the monodromy matrix. We refer to \cite{Bombardelli:2018jkj} for a description of the notation and of the method.

In particular, the crucial relations satisfied by the monodromy matrix (the so-called \textit{RTT relations}) is utilised to derive ``commutation" relations between its entries. The monodromy matrix is in fact regarded as a matrix in the auxiliary $0$ space with entries being operators on a chain of particles of length $N$, each particle having rapidity $\theta_i$. The vector notation $\vec{\theta}$ compactly denotes $\vec{\theta} = (\theta_1,...,\theta_N)$. Specifically,
\begin{eqnarray}
\label{eij}
&&{\cal{M}}(\theta_0|\vec{\theta}\, ) = E_{11} \otimes A(\theta_0|\vec{\theta}\, ) + E_{12} \otimes B(\theta_0|\vec{\theta}\, )+ E_{21} \otimes C(\theta_0|\vec{\theta}\, )+ E_{22} \otimes D(\theta_0|\vec{\theta}\, ),
\end{eqnarray}
where the matrix-unities $E_{ij}$ in (\ref{eij}) are taken in space $0$, and the operators $A,B,C,D$ act on the chain of $N$ particles.

Among the RTT relations, those which are needed to build the spectrum are the following:
\begin{eqnarray}
\label{rels}
&&A(\beta_1|\vec{\theta}\, )B(\beta_2|\vec{\theta}\, ) = X(\beta_1-\beta_2) B(\beta_1|\vec{\theta}\, )A(\beta_2|\vec{\theta}\, ) +Y(\beta_1-\beta_2) B(\beta_2|\vec{\theta}\, )A(\beta_1|\vec{\theta}\, ), \nonumber \\
&&D(\beta_1|\vec{\theta}\, )B(\beta_2|\vec{\theta}\, ) = X(\beta_1-\beta_2) B(\beta_1|\vec{\theta}\, )D(\beta_2|\vec{\theta}\, ) + Y(\beta_1-\beta_2) B(\beta_2|\vec{\theta}\, )D(\beta_1|\vec{\theta}\, ).
\end{eqnarray}
The coefficient functions $X,Y$ are now different from \cite{Bombardelli:2018jkj}, and read for the case at hand:
\begin{eqnarray}
\label{xy}
&&X(\beta) = \frac{a(\beta)}{b(\beta)} = \frac{2 i \, e^{\frac{\beta}{2}} \sin \frac{\pi}{k}}{e^\beta - 1}, \qquad  Y(\beta) = \frac{c(\beta)}{b(\beta)} = \frac{e^{-\frac{i \pi}{k}}(e^\beta - e^{\frac{2 i \pi}{k}})}{e^\beta - 1}.
\end{eqnarray}
The relations (\ref{rels}) allow us to postulate that a generic eigenstate of the transfer matrix 
\begin{equation}
{\cal{T}}(\theta_0|\vec{\theta}\, ) = A(\theta_0|\vec{\theta}\, ) - D(\theta_0|\vec{\theta}\, ),
\end{equation}
will be given by 
\begin{equation}
|\beta_1,...,\beta_M\rangle = B(\beta_1|\vec{\theta}\, )...B(\beta_M|\vec{\theta}\, )|0\rangle,
\end{equation}
provided one imposes a set of conditions on the rapidities. More precisely, one first notices that the pseudo-vacuum itself is an eigenstate of the transfer matrix (as can be proven using a recursive argument):
\begin{equation}
\Big[A(\theta_0|\vec{\theta}\, ) - D(\theta_0|\vec{\theta}\, )\Big] \, |0\rangle = \Big[1 - \prod_{i=1}^N b(\theta_0 - \theta_i)\Big] \, |0\rangle. 
\end{equation}
Then, using the relations (\ref{rels}) and making use of special relations between the coefficient functions, one manages to find that
\begin{eqnarray}
\label{use}
&&\Big[A(\theta_0|\vec{\theta}\, ) - D(\theta_0|\vec{\theta}\, )\Big] \, |\beta_1,...,\beta_M\rangle = \Lambda(\theta_0|\vec{\beta}|\vec{\theta}\, )\, |\beta_1,...,\beta_M\rangle \nonumber\\
&&\, \, \, + \sum_{j=1}^M X(\theta_0 - \beta_j)\Big[\prod_{k \neq j}^M Y(\beta_j - \beta_k)\Big] \, \Big[1 - \prod_{i=1}^N b(\beta_j - \theta_i)\Big]|\beta_1,...,\beta_{j-1}, \theta_0, \beta_{j+1},...,\beta_M\rangle,
\end{eqnarray}
with
\begin{equation}
\Lambda(\theta_0|\vec{\beta}|\vec{\theta}\, ) = \Big[\prod_{i=1}^M Y(\theta_0 - \beta_i)\Big] \Big[1 - \prod_{s=1}^N b(\theta_0 - \theta_s)\Big]
\end{equation}
and $\vec{\beta} = (\beta_1,...,\beta_M)$. We have verified this explicitly up to $M=3$ (with $N$ being kept generic and larger than $3$). The relation 
\begin{equation}
Y(\theta - \alpha) X(\theta - \beta) + X(\theta - \alpha) X(\alpha - \beta) = X(\theta - \beta) Y(\beta - \alpha)
\end{equation}
is especially useful in obtaining (\ref{use}).

At this point we can see that the state $|\beta_1,...,\beta_M \rangle$ is a transfer-matrix eigenstate if we kill the unwanted terms in the second line of (\ref{use}), which we can do by means of the \textit{auxiliary Bethe equations}
\begin{equation}
\label{auxi}
\prod_{i=1}^N b(\beta_j - \theta_i) = 1, \qquad j=1,...,M,
\end{equation}
with the function $b(\theta)$ given in (\ref{abc}).

The main set of Bethe equations (so-called \textit{momentum carrying}) are then obtained by imposing
\begin{equation}
e^{i e^{\theta_k} L} \Lambda(\theta_k|\vec{\beta}|\vec{\theta}\, ) = 1, \qquad k=1,...,N.
\end{equation}
Considering that we still need to take into account the dressing factor - which we denote as $\Phi$ - and that $b(0)=0$, we get as a momentum-carrying equation
\begin{equation}
\label{main}
e^{i e^{\theta_k} L} \prod_{j=1}^N \Phi(\theta_k - \theta_j)  \prod_{i=1}^M Y(\theta_k - \beta_i) = 1, \qquad k=1,...,N,
\end{equation}
with the function $Y(\theta)$ given in (\ref{xy}).

\section{TBA for relativistic limit of mixed-flux theory}
\label{sec6}

In this section we perform the TBA analysis, closely following section 5.2 in \cite{Bombardelli:2018jkj}. The first thing to notice by experimenting with the numerics, is that the solutions to the auxiliary Bethe equations (\ref{auxi}) still localise along two lines in the complex $\beta$-plane\footnote{The number of such solution is also what is expected to build the spectrum from the ABA as in appendix B.2.2 of \cite{Bombardelli:2018jkj}.}. However these two lines are not any longer at $\beta = z \pm i \frac{\pi}{2}$ with $z$ real, as in the pure R-R case of \cite{Bombardelli:2018jkj}, but instead they localise at 
\begin{equation}
\beta = z + i \pi \big(1 - \frac{1}{k}\big), \qquad \beta = z - i \frac{\pi}{k}, \qquad z \, \, \mbox{real}.
\end{equation}
This means that there will still be two separate kernels for the densities of auxiliary Bethe roots (\ref{auxi}), however they will be given by the following formulas:
\begin{eqnarray}
&&\phi_+ = \frac{1}{2 \pi i}\frac{d}{d\theta}\log b\Big(\theta + i \pi \big(1 - \frac{1}{k}\big )\Big), \qquad  \phi_- = \frac{1}{2 \pi i}\frac{d}{d\theta}\log b\Big(\theta - i \frac{\pi}{k}\Big),
\end{eqnarray}
with the function $b(\theta)$ given in (\ref{abc}).
This produces the kernels
\begin{equation}
\label{Theta}
\phi_\pm = -\frac{\sin \frac{\pi}{k}}{2\pi(\cos \frac{\pi}{k} \pm \cosh\theta)}.\nonumber
\end{equation}
The two kernels are not simply one the opposite of the other, as in \cite{Bombardelli:2018jkj}.

However, this is not the only difference with respect to the pure R-R case. The momentum-carrying Bethe equations have a different function controlling the contribution from the auxiliary roots, which is not simply the reciprocal of $b$. The function appearing in (\ref{main}) is the function $Y$ given in (\ref{xy}). This means that we need to compute two more kernels:
\begin{eqnarray}
&&\xi_+ = \frac{1}{2 \pi i}\frac{d}{d\theta}\log Y\Big(\theta + i \pi \big(1 - \frac{1}{k}\big )\Big), \qquad \xi_- = \frac{1}{2 \pi i}\frac{d}{d\theta}\log Y\Big(\theta - i \frac{\pi}{k}\Big),
\end{eqnarray}
which results in
\begin{equation}
\xi_\pm = -\frac{\sin \frac{\pi}{k}}{2\pi(\cos \frac{\pi}{k} \pm \cos(\frac{2\pi}{k}+i\theta)},
\end{equation}
with the same $\Theta$ variable as in (\ref{Theta}). The equations for the densities are
\begin{eqnarray}
\label{costo}
&&\rho_0^r(\theta) + \rho_0^h(\theta) = \frac{e^\theta}{2 \pi} + 2 \phi_0 * \rho_0^r - \sum_{n=1,3} (\xi_- * \rho_{-n}^r + \xi_+ * \rho_{+n}^r),\nonumber\\
&&\rho_{\pm n}^r(\beta) + \rho_{\pm n}^h(\beta) = \mp \phi_\pm * \rho_0^r, \qquad n=1,3,
\end{eqnarray}
where 
\begin{equation}
\phi_0 = \frac{1}{2 \pi i} \frac{d}{d \theta} \log \Phi_{LL}(\theta)\,.
\end{equation}
Minimising the free energy subject to the constraint~\eqref{costo} leads to the TBA equations:
\begin{eqnarray}
&&R \epsilon(\theta) = \epsilon_0(\theta) + 2 [\phi_0 * \log(1+e^{- \epsilon_0})](\theta) + \sum_{a=\pm, n=1,3} \bar{a} [\phi_a * \log(1+e^{- \epsilon_{a,n}})](\theta),\nonumber\\
&&\epsilon_{a,n}(\beta) = [\xi_a * \log(1+e^{- \epsilon_0})](\beta), \qquad a = \pm, \qquad n = 1,3,
\end{eqnarray}
where we have used the same notational conventions as in \cite{Bombardelli:2018jkj} for the pseudo-energies, $\epsilon(\theta) = e^\theta$, and the same trick of variable-change has been used to free the density-variations from the convolutions and put them in evidence, which is an essential step to obtain, for generic variations, the above TBA equations. We have also denoted $\bar{+}=-$ and $\bar{-}=+$.

If we compute the Witten index based on the above TBA, we just need to shift all the auxiliary pseudo-energies by $\pm i \pi$ in the L-functions. An exact solution of the TBA is given by $\epsilon_0 = \infty$ and $\epsilon_{a,n}=0$, so that Witten's index is zero. A little experimenting displays in fact that the integral of the kernels $\int_{-\infty}^\infty d\theta \, \bar{a} \phi_a(\theta)$ is a strictly positive quantity for $k>1$. Supersymmetry of the ground state is then preserved.   

\section{The relativistic limit for small $k$}
\label{sec7}

The WZW level $k$ enters as a simple parameter into the integrable S matrix construction of the NS-NS theory. As such, one is free to set it to small values like $k=1,2$ and consider the resulting integrable system in the same way as one does for other values of $k$. Since the relativistic limit captures the low-momentum infra-red physics of the gapless excitations, it should be already interesting to consider $k=1,2$ in the relativistic limit. Here we find an interesting subtlety. Setting $k=2$ in equation~\eqref{eq:RLLlim1} gives the S matrix
\begin{equation}\label{eq:RLLlimkis2}
  \begin{aligned}
    R_{LL} |\phi\rangle \otimes |\phi\rangle &= |\phi\rangle \otimes |\phi\rangle, \\
    R_{LL} |\phi\rangle \otimes |\psi\rangle &= -e^{i\varphi}\frac{ q_1-q_2}{q_1+q_2} |\phi\rangle \otimes |\psi\rangle + \frac{2\sqrt{q_1q_2} }{q_1+q_2} |\psi\rangle \otimes |\phi\rangle, \\
    R_{LL}|\psi\rangle \otimes |\phi\rangle &= \frac{2\sqrt{q_1q_2} }{q_1+q_2}|\phi\rangle \otimes |\psi\rangle +  e^{-i\varphi}\frac{q_1-q_2}{q_1+q_2} |\psi\rangle \otimes |\phi\rangle, \\
    R_{LL} |\psi\rangle \otimes |\psi\rangle &= - |\psi\rangle \otimes |\psi\rangle\,,
  \end{aligned}
\end{equation}
where $\varphi$ is a constant phase that is the relativistic limit of the (string) frame-factor, and can be set to zero by a simple redefinition of the two-excitation states.~\footnote{Frame factors, conventionally denoted by $\nu= (x^+_{L\,p}/x^-_{L\,p})^{1/2} = e^{\frac{i}{2} p}$, are of course present  for any $k$, and can be similarly removed. Nonethless, one needs to be particularly careful with them for $k=1,2$ when taking the relativistic limit. This is because upon shifting the momentum, in the relativistic limit they acquiring a factor $\nu \sim ( e^{-\frac{2i\pi}{k}} )^{1/2}$, which has a discontinuity at $k=2$.} The above S matrix can be also obtained by first setting $k=2$ and then taking the low-momentum relativistic limit. 

Turning to $k=1$, we see immediately that setting $k=1$ in equation~\eqref{eq:RLLlim1} gives a constant S matrix whose entries are $\pm 1$ and $0$. However, if one is more careful and first sets $k=1$ in the full non-relativistic S matrix and then takes the low-momentum limit one obtains the $k=2$ relativistic S matrix~\eqref{eq:RLLlimkis2}!~\footnote{This is because in a low-momentum expansion at generic values of $k$, higher-order momentum terms are divergent when one sets $k=1$.}. A similar analysis can be carried out for the dressing factors and tells us that minimal solutions to the relativistic limit of crossing equations are the same for $k=1$ and $k=2$, and can be obtained simply by setting $k=2$ in $\Phi_{LL}$ (see the discussion at the end of section \ref{sec4}) to obtain
\begin{equation}
\label{zamo}
\Phi_{LL}(\theta) = \prod_{\ell=1}^\infty \frac{\Gamma^2(\ell - \tau) \, \Gamma(\frac{1}{2} + \ell + \tau) \,\Gamma(- \frac{1}{2} + \ell + \tau)}{\Gamma^2(\ell + \tau) \, \Gamma(\frac{1}{2} + \ell - \tau) \,\Gamma(- \frac{1}{2} + \ell - \tau)},
\end{equation}
with $\tau = \frac{\theta}{2 \pi i}$.

The S matrix one obtains from the R-matrix~\eqref{eq:RLLlim1} by setting $k=2$ and dressing factor~\eqref{zamo} can be easily recognised as the minimal ${\cal{N}}=2$ S-matrix of Fendley-Intriligator \cite{FI}, namely the sine-Gordon S-matrix at the special value of the coupling  $\beta^2 = \frac{16 \pi}{3}$. Fendley and Intriligator used this S-matrix, coupled with a suitable mixed left-right scattering matrix, to describe a massless integrable flow between a UV $c=3$ CFT and an IR $c=1$ CFT, in the context of Landau-Ginzburg models. 

In our setting, the left-right scattering matrix is trivial, and the low-momentum gapless theory we are describing using a massless purely left-left and right-right S-matrix is a CFT~\cite{Zamol2}. This is exactly analogous to the CFT${}^{(0)}$ we introduced in~\cite{Bombardelli:2018jkj} to describe the low-energy gapless part of the spectrum of the $m=0$ modes in the pure R-R theory. As we have seen, in the NS-NS theory, excitations with $m=0,1,-1$ all give rise to such CFTs, and we will refer to them as CFT${}^{(0)}_m$. The $m=0$ case is identical to the R-R case~\cite{Bombardelli:2018jkj}, and the relativistic TBA was used there to show that the theory has central charge
\begin{equation}
c_{m=0}=6\,.
\end{equation}
In a forthcoming publication we will determine the central charge of CFT${}^{(0)}_{m=\pm 1}$ using the TBA found in Section~\ref{sec6} for generic values of $k$. At present however, we can use the fact that at $k=2,1$ the theory reduces precisely to the one considered in~\cite{FI}. As in~\cite{Bombardelli:2018jkj}, the central charge of interest to the relativistic limit considered in this paper is what is referred to in~\cite{FI} as the UV limit, so
\begin{equation}
c_{m=\pm 1}=3\,.
\end{equation}
In total then, we find a central charge of 12, and this should be interpreted as coming from the 8 transverse free bosons and their fermionic superpartners, as expected in the BMN limit of the NS-NS theory~\cite{Berenstein:2002jq}. This as an important consistency check of our relativistic TBA analysis. In particular, it confirms the absence of CDD factors in the dressing factors.~\footnote{We would like to thank Tim Hollowood for emphasizing this point to us.} As one goes away from the relativistic limit, by considering excitations with higher momentum, these free bosons and fermions will re-couple with one another and we will need the exact non-relativistic TBA involving the massive and massless excitations to find the spectrum.

\bigskip

\section{Conclusions}

In massive integrable theories, finite size corrections to the spectrum are exponentially suppressed in powers of
\begin{equation}
e^{-m L}\,.
\end{equation}
The degrees of freedom of integrable $\AdS_3/\CFT_2$ theories include excitations which, from the point of view of the worldsheet, are massless ($m=0$). As a result, including finite size corrections order-by-order in $\alpha'$ seems challenging in this setting~\cite{Abbott:2015pps}. On the other hand, in massless relativistic theories a TBA for finite size corrections is well-known~\cite{Zamol2}. In this paper, we use relativistic invariance to calculate finite size corrections due to massless excitations in integrable $\AdS_3/\CFT_2$ in two settings. Firstly, building on the hidden-relativistic invariance found in~\cite{Fontanella:2019baq}, we construct the exact in $\alpha'$ TBA for $\AdS_3$ R-R charge backgrounds, and show how it reduces to the low-energy TBA of~\cite{Bombardelli:2018jkj}. We use this exact TBA to show that the BMN vacuum does not receive finite size corrections. Secondly, we generalise the low-energy limit of~\cite{Bombardelli:2018jkj} to $\AdS_3$ backgrounds with NS-NS charge. In this limit, we find the dressing factors, ABA and TBA of the system, showing that NS-NS and hence mixed-charge $\AdS_3$ backgrounds' finite size corrections do not spoil integrability. 

It would be interesting to extend our NS-NS analysis away from the low-energy limit to obtain exact in $\alpha'$ expressions like the ones we found for the R-R case. Further, computing finite size corrections to excited states in both R-R and NS-NS theories is likely to be a useful step in developing a TBA and a Quantm Spectral Curve (QSC)~\cite{Gromov:2013pga,Cavaglia:2014exa,Gromov:2014caa,Bombardelli:2017vhk} for the combined massive-massless  theory.  Given the simple way in which moduli enter the integrable S matrix and BEs~\cite{OhlssonSax:2018hgc} incorporating them into the TBA and QSC should also be possible and would allow to compare with string field theory computations~\cite{Cho:2018nfn}. 

Following recent results on low-$k$ WZW strings in $\AdS_3$ backgrounds~\cite{Giribet:2018ada,Gaberdiel:2018rqv,Eberhardt:2018ouy}, we have shown that the relativistic limit of the low-$k$ integrable S matrices is well-defined. We found that in this limit the $k=1,2$ S matrices and dressing factors are both equal to  the minimal ${\cal{N}}=2$ S-matrix of Fendley-Intriligator \cite{FI}, equivalently the Sine-Gordon S-matrix at the special value of the coupling  $\beta^2 = \frac{16 \pi}{3}$. The equivalence of the $k=1$ and $k=2$ theories is a consequence of the relativistic limit: already at the next order in the expansion the two S matrices are different from one another, and so the
spectra of the two theories will be different from one another as is expected on general grounds~\cite{Giribet:2018ada,Gaberdiel:2018rqv,Eberhardt:2018ouy}. 

It is important to contrast our results with the ones of~\cite{Baggio,Dei}. In that context, the S matrix is assumed to be trivial apart from a CDD dressing factor for \textit{opposite} worldsheet chirality excitations. It would be interesting to see how the approach of~\cite{Baggio,Dei} compares with the one which we have developed in this paper, and whether the analysis of \cite{Baggio,Dei} can be reconciled with the general expectations~\cite{Zamol2} for an S-matrix description of a two-dimensional $\CFT$.

Finally, we observe that mixed-flux S matrices (both relativistic and exact) are well defined in the $\alg{psu}(1|1)^2$ sub-sector: indeed for the most part we wrote down explicit expressions just for such a sector. Viewed as a stand-alone object, these S matrices appear to be well-suited for describing mixed-flux $\AdS_2$ string backgrounds such as the $AdS_2\times\Sphere^2\times\Sphere^2\times \Torus^4$ background considered in~\cite{Wulff:2014kja}, whose M-theory origin comes from~\cite{Gauntlett:1997pk,Boonstra:1998yu}. In light of this it would be very interesting to investigate the potential integrable structure of this string theory background.~\footnote{We are grateful to the anonymous referee for reminding us of the existence of this background and the paper~\cite{Wulff:2014kja}.}

\section{Acknowledgments}
We very much thank Diego Bombardelli for discussions, collaboration on the initial stages of the project and continuing assistance. We very much thank Ben Hoare, Tim Hollowood and Arkady Tseytlin for reading the manuscript and for very useful comments and discussions. We thank Alessandra Cagnazzo, Scott Collier, Andrea Dei, Marius de Leeuw, Jan Gutowski, Sergey Frolov, Tim Hollowood, Tristan McLoughlin, Antonio Pittelli, Anton Pribytok, Ana Retore, Paul Ryan, Xi Yin and Kostya Zarembo for discussions. We thank Lorenzo Bianchi for comments on the comparison with perturbation theory. B. S. acknowledges funding support from an STFC Consolidated Grant `Theoretical Physics at City University" ST/P000797/1. This research was supported in part by Perimeter Institute for Theoretical Physics. Research at Perimeter Institute is supported by the Government of Canada through the Department of Innovation, Science, and Economic Development, and by the Province of Ontario through the Ministry of Research and Innovation. A. T. thanks the STFC for support under the Consolidated Grant project nr. ST/L000490/1. This work was supported by the ERC advanced grant No 341222. 
A. F. acknowledges the support of the \emph{Angelo Della Riccia Foundation} Fellowship. 
No data beyond those presented and cited in this work are needed to validate this study.

\setcounter{section}{0}
\setcounter{subsection}{0}

\begin{appendices}

\section{BES coefficients and relativistic invariance}
\label{App:BES}
The coefficients $c_{r,s} (g)$ for the BES dressing factor in the large $x, y$ expansion (\ref{dressing_expansion}) are
\begin{equation}
\label{BES_coeff}
c_{r,s} (g) = 2 \sin\big[\frac{\pi}{2} (r-s) \big] \int_0^{\infty} dt \, \frac{J_r(2gt) J_s(2gt)}{t(e^t -1)} \ , 
\end{equation}
where $g$ is the 't Hooft coupling, and $J_n$ are Bessel functions.
We want to show that the BES coefficients (\ref{BES_coeff}) do not satisfy the relativistic invariance condition (\ref{rel_inv_cond}). 
The relativistic invariance condition for the BES coefficients is the following 
\begin{eqnarray}
\label{BES_rel_inv}
\notag
&&2(r+1) \cos \big[ \frac{\pi}{2} (r-s) \big] \int_0^{\infty} dt \, \frac{J_{r+1} (2gt) J_s (2gt)}{t (e^t - 1)} \\
\notag
&&+ 2(r-1) \cos \big[ \frac{\pi}{2} (r-s) \big] \int_0^{\infty} dt \, \frac{J_{r-1} (2gt) J_s (2gt)}{t (e^t - 1)} \\
\notag
&&- 2(s+1) \cos \big[ \frac{\pi}{2} (r-s) \big] \int_0^{\infty} dt \, \frac{J_{r} (2gt) J_{s+1} (2gt)}{t (e^t - 1)} \\
&&- 2(s-1) \cos \big[ \frac{\pi}{2} (r-s) \big] \int_0^{\infty} dt \, \frac{J_{r} (2gt) J_{s-1} (2gt)}{t (e^t - 1)} = 0 
\end{eqnarray}
The above equation is always satisfied when $r - s = 2 k + 1$, for $k \in \mathbb{Z}$, and when $r = s$. However, we expect that it is not satisfied when $r - s = 2k$, for $k \neq 0$. Indeed we considered the cases $(r, s) = (3, 1), (5,1), (18, 10)$ and we numerically computed the l.h.s of (\ref{BES_rel_inv}) for values of the coupling $g = 0.1, 0.2 , ... , 1$ (increment of $0.1$), and we found that the l.h.s is always \emph{not} zero. Since equation (\ref{rel_inv_cond}) must hold for any values of $r,s$ in order for a given dressing factor to be relativistic invariant, this shows that the BES dressing factor is not relativistic invariant.

\end{appendices}

\end{document}